\documentclass[prd,aps,preprint,showpacs,preprintnumbers,amsmath,amssymb,tighten,nofootinbib,12pt]{revtex4}

\usepackage{graphicx}
\usepackage{dcolumn}
\usepackage{bm}

\begin{document}


\title{Generic Friedberg-Lee Symmetry of Dirac Neutrinos
\vspace{0.3cm}}

\author{{\bf Shu Luo}}
\email{luoshu@mail.ihep.ac.cn}

\author{{\bf Zhi-zhong Xing}}
\email{xingzz@mail.ihep.ac.cn}

\affiliation{Institute of High Energy Physics, Chinese Academy of
Sciences, Beijing 100049, China}

\author{{\bf Xin Li}}

\affiliation{Department of Physics, Wuhan University, Wuhan 430072,
China}

\begin{abstract}
We write out the generic Dirac neutrino mass operator which
possesses the Friedberg-Lee (FL) symmetry and find that its
corresponding neutrino mass matrix is asymmetric. Following a
simple way to break the FL symmetry, we calculate the neutrino
mass eigenvalues and show that the resultant neutrino mixing
pattern is nearly tri-bimaximal. Imposing the Hermitian condition
on the neutrino mass matrix, we also show that the simplified
ansatz is consistent with current experimental data and favors the
normal neutrino mass hierarchy.
\end{abstract}

\pacs{PACS number(s): 14.60.Lm, 14.60.Pq, 95.85.Ry}

\maketitle

\framebox{\large\bf 1} ~ Recent solar \cite{SNO}, atmospheric
\cite{SK}, reactor \cite{KM} and accelerator \cite{K2K} neutrino
experiments have provided us with very convincing evidence that
neutrinos are slightly massive and lepton flavors are
significantly mixed. The flavor mixing of three lepton families
can be described by a $3\times 3$ unitary matrix $U$ \cite{MNS},
which is usually parameterized as \cite{PDG}:
\begin{eqnarray}
U & = & \left ( \begin{matrix} c^{}_{12}c^{}_{13} & s^{}_{12}c
^{}_{13} & s^{}_{13} e^{-i\delta} \cr -s^{}_{12}c^{}_{23}
-c^{}_{12}s^{}_{23}s^{}_{13} e^{i\delta} & c^{}_{12}c^{}_{23}
-s^{}_{12}s^{}_{23}s^{}_{13} e^{i\delta} & s^{}_{23}c^{}_{13} \cr
s^{}_{12}s^{}_{23} -c^{}_{12}c^{}_{23}s^{}_{13} e^{i\delta} &
-c^{}_{12}s^{}_{23} -s^{}_{12}c^{}_{23}s^{}_{13} e^{i\delta} &
c^{}_{23}c^{}_{13} \end{matrix} \right)  \; ,
\end{eqnarray}
where $c^{}_{ij} \equiv \cos\theta_{ij}$ and $s^{}_{ij} \equiv
\sin\theta_{ij}$ (for $ij=12,13$ and $23$), and $\delta$ is the
CP-violating phase. If neutrinos are Majorana particles, $U$
should contain two more CP-violating phases, which are referred to
as the Majorana phases and have nothing to do with neutrino
oscillations. The latest global analysis of current neutrino
oscillation data yields $30.9^\circ \leq \theta^{}_{12} \leq
37.8^\circ$, $35.1^\circ \leq \theta^{}_{23} \leq 53.4^\circ$ and
$0^\circ \leq \theta^{}_{13} < 12.4^\circ$ with $3 \sigma$
uncertainty \cite{Fogli}, but the phase $\delta$ remains entirely
unconstrained. While the absolute mass scale of three neutrinos is
not yet fixed, their two mass-squared differences have already
been determined to a quite good degree of accuracy \cite{Fogli}:
$\Delta m^2_{21} \equiv m^2_2 - m^2_1 = (7.14 \cdot\cdot\cdot
8.19) \times 10^{-5} ~{\rm eV}^2$ and $\Delta m^2_{32} \equiv
m^2_3 - m^2_2 = \pm (2.06 \cdot\cdot\cdot 2.81) \times 10^{-3}
~{\rm eV}^2$ with $3 \sigma$ uncertainty.

Many theoretical and phenomenological attempts have been made to
interpret the smallness of three neutrino masses and the largeness
of two neutrino mixing angles \cite{Review}. Among them, the
flavor symmetry approach is in particular simple and predictive. A
new and intriguing flavor symmetry is the one proposed by
Friedberg and Lee (FL) \cite{FL}. In the basis where the mass
eigenstates of three charged leptons are identified with their
flavor eigenstates, the Dirac neutrino mass operator can be
written as
\begin{eqnarray}
{\cal L}^{}_{\rm FL} & = & \sum_{\alpha, \beta} Y^{}_{\alpha\beta}
\left ( \overline{\nu}^{}_{\alpha} - \overline{\nu}^{}_{\beta}
\right ) \left ( \nu^{}_{\alpha} - \nu^{}_{\beta} \right ) \; ,
\end{eqnarray}
where $\alpha$ and $\beta$ run over $e$, $\mu$ and $\tau$. The FL
symmetry means that ${\cal L}^{}_{\rm FL}$ is invariant under the
translational transformations $\nu^{}_e \rightarrow \nu^{}_e + z$,
$\nu^{}_\mu \rightarrow \nu^{}_\mu + z$ and $\nu^{}_\tau
\rightarrow \nu^{}_\tau + z$, where $z$ is a constant element of
the Grassmann algebra independent of space and time \cite{FL}. The
corresponding neutrino mass matrix is a symmetric matrix,
\begin{eqnarray}
M^{}_{\rm FL} & = & \left ( \begin{matrix} b + c & - b & - c \cr - b
& a + b & - a \cr - c & - a & a + c
\end{matrix} ~ \right ) \; ,
\end{eqnarray}
where $a = Y^{}_{\mu\tau} + Y^{}_{\tau\mu}$, $b = Y^{}_{e\mu} +
Y^{}_{\mu e}$ and $c = Y^{}_{\tau e} + Y^{}_{e\tau}$. Note that
the determinant of $M^{}_{\rm FL}$ is vanishing (i.e., ${\rm
Det}(M^{}_{\rm FL}) =0$), and thus one of the neutrinos must be
massless. One may explicitly break the FL symmetry of ${\cal
L}^{}_{\rm FL}$ to make realistic predictions for both neutrino
masses and flavor mixing angles. So far some interesting works
have been done to apply the FL symmetry to the Majorana neutrino
mass operator \cite{XZZ,LX,Xing07}, to combine the FL symmetry
with the seesaw mechanism \cite{Jarlskog,Chao}, to extend the FL
symmetry to the quark sector \cite{FL2,Ren}, and to generalize the
FL symmetry in a specific model containing some scalar fields
\cite{Liao}.

Here we notice that ${\cal L}^{}_{\rm FL}$ in Eq. (3) is not the
most generic mass operator of Dirac neutrinos which obeys the FL
symmetry. The Dirac neutrino mass operator
\begin{eqnarray}
{\cal L}^\prime_{\rm FL} & = & \sum_{\alpha, \beta} \sum_{\alpha',
\beta'} Y^{\alpha\beta}_{\alpha'\beta'} \left (
\overline{\nu}^{}_{\alpha} - \overline{\nu}^{}_{\beta} \right )
\left ( \nu^{}_{\alpha'} - \nu^{}_{\beta'} \right ) \; ,
\end{eqnarray}
where the Greek superscripts and subscripts run over $e$, $\mu$ and
$\tau$, is more general than ${\cal L}^{}_{\rm FL}$ and also
invariant under the translational transformations $\nu^{}_e
\rightarrow \nu^{}_e + z$, $\nu^{}_\mu \rightarrow \nu^{}_\mu + z$
and $\nu^{}_\tau \rightarrow \nu^{}_\tau + z$. Its corresponding
neutrino mass matrix $M^\prime_{\rm FL}$ takes the form
\begin{eqnarray}
M^\prime_{\rm FL} & = & \left ( \begin{matrix} ~ B + C & - B - D & -
C + D \cr - B + D & ~ A + B & - A - D \cr - C - D & - A + D & ~ A +
C \end{matrix} ~ \right ) \; ,
\end{eqnarray}
where
\small
\begin{eqnarray}
A & = & \frac{1}{2} \left [ - \left ( Y^{\tau \mu}_{\mu e} + Y^{\mu
\tau}_{e \mu} - Y^{\tau \mu}_{e \mu} - Y^{\mu \tau}_{\mu e} \right)
+ \left ( Y^{\mu e}_{e \tau} + Y^{e \mu}_{\tau e} - Y^{\mu e}_{\tau
e} - Y^{e \mu}_{e \tau} \right) - \left ( Y^{e \tau}_{\tau \mu} +
Y^{\tau e}_{\mu \tau} - Y^{e \tau}_{\mu \tau} - Y^{\tau e}_{\tau
\mu} \right) \right. \nonumber
\\ & & \left. - \left ( Y^{\tau \mu}_{e \tau} + Y^{\mu \tau}_{\tau e} -
Y^{\tau \mu}_{\tau e} - Y^{\mu \tau}_{e \tau} \right) - \left (
Y^{\mu e}_{\tau \mu} + Y^{e \mu}_{\mu \tau} - Y^{\mu e}_{\mu \tau} -
Y^{e \mu}_{\tau \mu} \right) + \left ( Y^{e \tau}_{\mu e} + Y^{\tau
e}_{e \mu} - Y^{e \tau}_{e \mu} - Y^{\tau e}_{\mu e} \right) \right
] \nonumber\\
& & +\left ( Y^{\tau \mu}_{\tau \mu} + Y^{\mu \tau}_{\mu \tau} -
Y^{\tau \mu}_{\mu \tau} - Y^{\mu \tau}_{\tau \mu} \right) \; ;
\nonumber \\
B & = & \frac{1}{2} \left [ - \left ( Y^{\tau \mu}_{\mu e} + Y^{\mu
\tau}_{e \mu} - Y^{\tau \mu}_{e \mu} - Y^{\mu \tau}_{\mu e} \right)
- \left ( Y^{\mu e}_{e \tau} + Y^{e \mu}_{\tau e} - Y^{\mu e}_{\tau
e} - Y^{e \mu}_{e \tau} \right) + \left ( Y^{e \tau}_{\tau \mu} +
Y^{\tau e}_{\mu \tau} - Y^{e \tau}_{\mu \tau} - Y^{\tau e}_{\tau
\mu} \right) \right. \nonumber
\\ & & \left. + \left (
Y^{\tau \mu}_{e \tau} + Y^{\mu \tau}_{\tau e} - Y^{\tau \mu}_{\tau
e} - Y^{\mu \tau}_{e \tau} \right) - \left ( Y^{\mu e}_{\tau \mu} +
Y^{e \mu}_{\mu \tau} - Y^{\mu e}_{\mu \tau} - Y^{e \mu}_{\tau \mu}
\right) - \left ( Y^{e \tau}_{\mu e} + Y^{\tau e}_{e \mu} - Y^{e
\tau}_{e \mu} - Y^{\tau e}_{\mu e} \right) \right ] \nonumber\\
& & + \left ( Y^{\mu e}_{\mu e} + Y^{e \mu}_{e \mu} - Y^{\mu e}_{e
\mu} - Y^{e \mu}_{\mu e} \right ) \; ;
\nonumber \\
C & = & \frac{1}{2} \left [ \left ( Y^{\tau \mu}_{\mu e} + Y^{\mu
\tau}_{e \mu} - Y^{\tau \mu}_{e \mu} - Y^{\mu \tau}_{\mu e} \right)
- \left ( Y^{\mu e}_{e \tau} + Y^{e \mu}_{\tau e} - Y^{\mu e}_{\tau
e} - Y^{e \mu}_{e \tau} \right) - \left ( Y^{e \tau}_{\tau \mu} +
Y^{\tau e}_{\mu \tau} - Y^{e \tau}_{\mu \tau} - Y^{\tau e}_{\tau
\mu} \right) \right. \nonumber
\\ & & \left. - \left (
Y^{\tau \mu}_{e \tau} + Y^{\mu \tau}_{\tau e} - Y^{\tau \mu}_{\tau
e} - Y^{\mu \tau}_{e \tau} \right) + \left ( Y^{\mu e}_{\tau \mu} +
Y^{e \mu}_{\mu \tau} - Y^{\mu e}_{\mu \tau} - Y^{e \mu}_{\tau \mu}
\right) - \left ( Y^{e \tau}_{\mu e} + Y^{\tau e}_{e \mu} - Y^{e
\tau}_{e \mu} - Y^{\tau e}_{\mu e} \right) \right ] \nonumber\\
& & + \left ( Y^{e \tau}_{e \tau} + Y^{\tau e}_{\tau e} - Y^{e
\tau}_{\tau e} - Y^{\tau e}_{e \tau} \right) \; ;
\nonumber \\
D & = & \frac{1}{2} \left [ \left ( Y^{\tau \mu}_{\mu e} + Y^{\mu
\tau}_{e \mu} - Y^{\tau \mu}_{e \mu} - Y^{\mu \tau}_{\mu e} \right)
+ \left ( Y^{\mu e}_{e \tau} + Y^{e \mu}_{\tau e} - Y^{\mu e}_{\tau
e} - Y^{e \mu}_{e \tau} \right) + \left ( Y^{e \tau}_{\tau \mu} +
Y^{\tau e}_{\mu \tau} - Y^{e \tau}_{\mu \tau} - Y^{\tau e}_{\tau
\mu} \right) \right. \nonumber
\\ & & \left. - \left ( Y^{\tau \mu}_{e \tau} + Y^{\mu \tau}_{\tau e}
- Y^{\tau \mu}_{\tau e} - Y^{\mu \tau}_{e \tau} \right) - \left (
Y^{\mu e}_{\tau \mu} + Y^{e \mu}_{\mu \tau} - Y^{\mu e}_{\mu \tau} -
Y^{e \mu}_{\tau \mu} \right) - \left ( Y^{e \tau}_{\mu e} + Y^{\tau
e}_{e \mu} - Y^{e \tau}_{e \mu} - Y^{\tau e}_{\mu e} \right) \right
] . ~~~~
\end{eqnarray}
\normalsize
We see that $M^\prime_{\rm FL}$ is an asymmetric matrix
and its asymmetry is characterized by non-vanishing $D$. Given $D =
0$, $M^\prime_{\rm FL}$ turns out to be equivalent to $M^{}_{\rm
FL}$.

Based on the above observation, we are going to focus our interest
on the phenomenological implications of ${\cal L}^\prime_{\rm FL}$
for Dirac neutrinos. We shall follow a simple way to break the FL
symmetry of ${\cal L}^\prime_{\rm FL}$ and obtain the neutrino
mass matrix $M^{}_\nu = M^\prime_{\rm FL} + m^{}_0 {\bf 1}$ with
$\bf 1$ being the identity matrix. Then we shall show that a
nearly tri-bimaximal neutrino mixing pattern, which is favored by
current neutrino oscillation data, can always be obtained from
$M^{}_\nu$. A simpler and Hermitian form of $M^{}_\nu$ will also
be discussed in detail.

It is worth remarking that the nature of massive neutrinos remains
unclear, although most theorists believe that they should be
Majorana particles. However, there {\it do} exist some interesting
models in the literature \cite{Dirac}, where massive neutrinos are
treated as Dirac fermions. Before the nature of neutrinos is
experimentally identified, we feel that it makes sense to study the
phenomenology of both Dirac and Majorana neutrinos.

\vspace{0.3cm}

\framebox{\large\bf 2} ~ Although $M^\prime_{\rm FL}$ in Eq. (5) is
asymmetric, it is easy to verify that its determinant vanishes as
$M^{}_{\rm FL}$ does. Hence one of the mass eigenvalues of
$M^\prime_{\rm FL}$ must be zero. To generate non-vanishing masses
for all the three neutrinos, here we follow Ref. \cite{FL} to break
the FL symmetry of ${\cal L}^\prime_{\rm FL}$:
\begin{eqnarray}
{\cal L}^{}_\nu & = & {\cal L}^\prime_{\rm FL} + m^{}_{0}
\sum_\alpha \overline{\nu}^{}_\alpha \nu^{}_\alpha \; ,
\end{eqnarray}
where $m^{}_0$ is in general a complex parameter, and $\alpha$ runs
over $e$, $\mu$ and $\tau$. Corresponding to ${\cal L}^{}_\nu$, the
Dirac neutrino mass matrix reads
\begin{eqnarray}
M^{}_{\nu} & = & M^\prime_{\rm FL} + m^{}_0 {\bf 1} \; = \; \left (
\begin{matrix} ~ B + C & - B - D & - C + D \cr - B + D & ~ A + B & -
A - D \cr - C - D & - A + D & ~ A + C
\end{matrix} ~ \right ) +
m^{}_{0} \left ( \begin{matrix} ~ 1 ~ & 0 & 0 \cr 0 & ~ 1 ~ & 0 \cr
0 & 0 & ~ 1 ~ \end{matrix} \right ) \; .
\end{eqnarray}
We see that ${\cal L}^{}_\nu$ or $M^{}_\nu$ can possess the exact
$\mu$-$\tau$ symmetry only when both $B=C$ and $D=0$ are satisfied.
To derive the neutrino mass spectrum and the flavor mixing pattern
from $M^{}_\nu$, we consider the following unitary transformation:
\begin{eqnarray}
U^{\dagger}_{} M^{}_\nu M^\dagger_\nu U^{}_{} & = & \left (
\begin{matrix} ~ m^2_1 ~ & 0 & 0 \cr 0 & ~ m^2_2 ~ & 0 \cr
0 & 0 & ~ m^2_3 ~ \end{matrix} \right ) \; ,
\end{eqnarray}
where $m^{}_{i}$ (for $i = 1, 2, 3$) stand for three neutrino
masses. Because we have taken the basis in which the mass and flavor
eigenstates of three charged leptons are identical, the unitary
matrix $U$ in Eq. (9) is just the neutrino mixing matrix linking the
neutrino mass eigenstates $(\nu^{}_1, \nu^{}_2, \nu^{}_3)$ to the
neutrino flavor eigenstates $(\nu^{}_e, \nu^{}_\mu, \nu^{}_\tau)$.

A salient feature of $M^{}_\nu$ is that the sum of three elements
in its any row or column equals $m^{}_0$, implying that one of its
three eigenvalues must be $m^{}_0$. For this reason, the unitary
transformation $U$ used to diagonalize the Hermitian matrix
$M^{}_{\nu}M^\dagger_\nu$ must have an eigenvector which contains
three equal components $1 / \sqrt{3}$. It is then possible to
express $U$ as a production of the tri-bimaximal mixing matrix
$U^{}_0$ \cite{TB} and a complex rotation matrix $U^{}_\theta$ in
the (1,3) plane:
\begin{eqnarray}
U & = & U^{}_{0} \otimes U^{}_{\theta} \; = \; \left (
\begin{matrix} \displaystyle ~ \frac{2}{\sqrt{6}} &
\displaystyle ~ \frac{1}{\sqrt{3}} & ~ 0 \cr \displaystyle -
\frac{1}{\sqrt{6}} & \displaystyle ~ \frac{1}{\sqrt{3}} &
\displaystyle ~ \frac{1}{\sqrt{2}} \cr \displaystyle -
\frac{1}{\sqrt{6}} & \displaystyle ~ \frac{1}{\sqrt{3}} &
\displaystyle - \frac{1}{\sqrt{2}}
\end{matrix} ~ \right ) \otimes \left ( \begin{matrix} \cos\theta & 0
& ~ \sin\theta ~ e^{- i \delta}_{}  \cr 0 & 1 & 0 \cr - \sin\theta ~
e^{i \delta}_{} & ~0~ & \cos\theta \end{matrix} \right ) \; ,
\end{eqnarray}
in which $\delta$ signifies CP violation and is equivalent to the
one defined in Eq. (1). After a straightforward calculation, we
obtain
\begin{eqnarray}
\delta & = & - {\rm arg} \left ( T^{}_{13} \right ) \; ,
\nonumber \\
\theta & = & \frac{1}{2} \arctan \left(\frac{2
|T^{}_{13}|}{T^{}_{33} - T^{}_{11}}\right) \; ,
\end{eqnarray}
where
\begin{eqnarray}
T^{}_{11} & = & 3 \left ( |B|^{2}_{} + |C|^{2}_{} + {\rm
Re}[B^{*}_{} C] + |D|^{2}_{} - {\rm Re}[\left ( C - B \right )
D^{*}_{}] \right ) + 3 {\rm Re}[\left ( B + C \right ) m^{*}_{0}] +
|m^{}_{0}|^{2}_{} \; ,
\nonumber \\
T^{}_{33} & = & |B|^{2}_{} + |C|^{2}_{} - {\rm Re}[B^{*}_{} C] + 4
|A|^{2}_{} + 2 {\rm Re}[\left ( B + C \right ) A^{*}_{}] + 3 {\rm
Re}[\left ( C - B \right ) D^{*}_{}]
\nonumber\\
& & + 3 |D|^{2}_{} + 4 {\rm Re}[A m^{*}_{0}] + {\rm Re}[\left ( B +
C \right ) m^{*}_{0}] + |m^{}_{0}|^{2}_{} \; ,
\nonumber \\
T^{}_{13} & = & \sqrt{3} \left ( |C|^{2}_{} - |B|^{2}_{} - i {\rm
Im}[B^{*}_{} C] \right ) + \sqrt{3} {\rm Re}[\left ( B + C \right )
D^{*}_{}] + 2 \sqrt{3} i {\rm Im}[\left ( B + C \right ) D^{*}_{}]
\nonumber\\
& & + \sqrt{3} \left ( C -B \right ) A^{*}_{} - 2 \sqrt{3} A^{*}_{}
D + \sqrt{3} {\rm Re}[\left( C - B \right ) m^{*}_{0}] - 2 \sqrt{3}
i {\rm Im}[D m^{*}_{0}] \; .
\end{eqnarray}
Furthermore, three mass eigenvalues of $M^{}_{\nu}$ are found to be
\begin{eqnarray}
m^{}_{1} & = & \sqrt{\frac{1}{2} \left ( T^{}_{11} + T^{}_{33}
\right ) - \frac{1}{2} \left ( T^{}_{33} - T^{}_{11} \right )
\cos2\theta - |T^{}_{13}| \sin2\theta} \;\; ,
\nonumber \\
m^{}_{2} & = & |m^{}_{0}| \; ,
\nonumber \\
m^{}_{3} & = & \sqrt{\frac{1}{2} \left ( T^{}_{11} + T^{}_{33}
\right ) + \frac{1}{2} \left ( T^{}_{33} - T^{}_{11} \right )
\cos2\theta + |T^{}_{13}| \sin2\theta} \;\; .
\end{eqnarray}
Comparing between Eqs. (1) and (10), one may easily arrive at the
analytical results of three neutrino mixing angles:
\begin{eqnarray}
\sin\theta^{}_{12} & = & \frac{1}{\sqrt{2 + \cos2\theta}} \; ,
\nonumber \\
\sin\theta^{}_{23} & = & \frac{\sqrt{2 + \cos2\theta - \sqrt{3}
\sin2\theta \cos\delta}}{\sqrt{2 \left ( 2 + \cos2\theta \right )}}
\; ,
\nonumber \\
\sin\theta^{}_{13} & = & \frac{2}{\sqrt{6}} |\sin\theta| \; .
\end{eqnarray}
In addition, we find that the Jarlskog invariant of leptonic CP
violation \cite{J} is given by ${\cal J} = \sin2\theta \sin\delta
/(6\sqrt{3})$ in this phenomenological scenario of Dirac neutrino
mixing.

Note that $A$, $B$, $C$, $D$ and $m^{}_0$ in $M^{}_\nu$ can all be
complex parameters. Hence it is always possible to find some
proper parameter space in which the neutrino mass spectrum
obtained in Eq. (13) and the neutrino mixing pattern obtained in
Eq. (14) are both compatible with current neutrino oscillation
data. In particular, no fine-tuning is needed to make $U$
consistent with the solar and atmospheric neutrino experiments
because $U$ itself is a nearly tri-bimaximal mixing pattern with
small $\theta$. Instead of carrying out a numerical analysis of
$m^{}_i$ and $\theta^{}_{ij}$ changing with those model
parameters, we shall look at a more specific scenario with
$M^{}_{\nu}$ being Hermitian in the following.

\vspace{0.3cm}

\framebox{\large\bf 3} ~ Given the asymmetric form of $M^{}_{\nu}$
in Eq. (8), the Hermitian relation $M^\dagger_\nu = M^{}_\nu$ can
be achieved if and only if $A$, $B$, $C$ and $m^{}_{0}$ are all
real and $D$ is purely imaginary (i.e., $D^* = -D$). Let us define
$D = i D'$ and rewrite $M^{}_{\nu}$ as
\begin{eqnarray}
M^{}_{\nu} & = & \left ( \begin{matrix} ~ B + C & - B - i D' & - C
+ i D' \cr - B + i D' & ~ A + B & - A - i D' \cr - C - i D' & - A
+ i D' & ~ A + C \end{matrix} ~ \right ) + m^{}_{0} \left (
\begin{matrix} ~ 1 ~ & 0 & 0 \cr 0 & ~ 1 ~ & 0 \cr 0 & 0 & ~ 1 ~
\end{matrix} \right ) \; ,
\end{eqnarray}
where $A$, $B$, $C$, $D'$ and $m^{}_{0}$ are all real. Now
$M^{}_\nu$ is Hermitian and only contains five free parameters. We
are going to show that this interesting texture of $M^{}_\nu$ is
actually compatible with current neutrino oscillation data.

With the help of Eqs. (11)---(14), it is straightforward to obtain
three neutrino masses and three flavor mixing angles from
Hermitian $M^{}_\nu$ given in Eq. (15). First,
\begin{eqnarray}
m^{}_{1} & = & \left | \left ( A + B + C + m^{}_{0} \right ) \mp
\sqrt{\left ( A^{2}_{} + B^{2}_{} + C^{2}_{} \right ) - \left ( A
B + B C + C A \right ) + 3 D'^{2}_{}} \right | \; ,
\nonumber \\
m^{}_{2} & = & |m^{}_{0}| \; ,
\nonumber \\
m^{}_{3} & = & \left | \left ( A + B + C + m^{}_{0} \right ) \pm
\sqrt{\left ( A^{2}_{} + B^{2}_{} + C^{2}_{} \right ) - \left ( A
B + B C + C A \right ) + 3 D'^{2}_{}} \right | \; .
\end{eqnarray}
Second,
\begin{eqnarray}
\sin\theta^{}_{12} & = & \frac{1}{\sqrt{2 + \cos2\theta}} \; ,
\nonumber \\
\sin\theta^{}_{23} & = & \frac{\sqrt{2 + \cos2\theta - \sqrt{3}
\sin2\theta \cos\delta}}{\sqrt{2 \left ( 2 + \cos2\theta \right )}}
\; ,
\nonumber \\
\sin\theta^{}_{13} & = & \frac{2}{\sqrt{6}} |\sin\theta| \; ,
\end{eqnarray}
where
\begin{eqnarray}
\delta & = & \arctan\left(\frac{2 D'}{C - B}\right) \; ,
\nonumber \\
\theta & = & \frac{1}{2} \arctan\left(\frac{\sqrt{3 \left [ \left
( C - B \right )^{2}_{} + 4 D'^{2}_{} \right ]}}{2 A - B -
C}\right) \; .
\end{eqnarray}
Note that $\delta$ is just the CP-violating phase of $U$, and
$\theta$ has been restricted to the range $- \pi/4 \leq \theta
\leq \pi/4$. Note also that $\theta >0$ and $\theta <0$ correspond
to the options of ``$\mp$" signs in the expression of $m^{}_1$ (or
the options of ``$\pm$" signs in the expression of $m^{}_3$) in
Eq. (16). Taking account of current experimental constraints on
three mixing angles \cite{Fogli}, we obtain $|\theta| <
18^{\circ}$. The smallness of $|\theta|$ implies that $U$ is a
nearly tri-bimaximal neutrino mixing pattern.

If both $B = C$ and $D' = 0$ hold, then $M^{}_{\nu}$ possesses the
exact $\mu$-$\tau$ symmetry which gives rise to the exact
tri-bimaximal neutrino mixing pattern $U^{}_0$ (i.e.,
$\theta^{}_{12} = \arctan(1/\sqrt{2}) \approx 35.3^{\circ}_{}$,
$\theta^{}_{13} = 0^\circ$ and $\theta^{}_{23} = 45^{\circ}_{}$).
There are two simpler ways to produce the deviation of $U$ from
$U^{}_0$:
\begin{enumerate}
\item $B \neq C$ and $D' = 0$. In this special case, we have
$\theta^{}_{13} \neq 0^\circ$ and $\theta^{}_{23} \neq 45^{\circ}$
together with $\delta = 0^\circ$ (CP conservation).

\item $B = C$ and $D' \neq 0$. In this special case, we have
$\delta = \pm\pi/2$ (CP violation), $\theta^{}_{23} = 45^{\circ}$
and $\theta^{}_{13} \neq 0^\circ$.
\end{enumerate}
The second possibility is more interesting in the sense that $|{\cal
J}| = \sin2\theta/(6\sqrt{3})$ can be as large as a few percent for
$|\theta| \geq 3^\circ$ and may lead to observable CP-violating
effects in long-baseline neutrino oscillations.

To illustrate, let us carry out a simple numerical analysis of the
parameter space of Hermitian $M^{}_\nu$ by using current neutrino
oscillation data on $(\Delta m^2_{21}, \Delta m^2_{32})$ and
$(\theta^{}_{12}, \theta^{}_{13}, \theta^{}_{23})$ as the inputs.
Without loss of generality, we assume $m^{}_{0} >0$. Our numerical
results indicate that only the normal neutrino mass hierarchy
(i.e., $\Delta m^2_{32} >0$) is favored in this Hermitian ansatz.
The allowed regions of $A$, $B$, $C$, $D'$ and $m^{}_{0}$ are
shown in Fig. 1, where $m^{}_{0} \lesssim 0.2$ eV has been taken
as a generous upper bound on the absolute neutrino mass scale
\cite{WMAP}. Because of $m^{}_{0} = m^{}_{2}$, the lower bound of
$m^{}_{0}$ is $m^{}_0 > \sqrt{\Delta m^{2}_{21}} \approx 0.09$ eV
as one can see from Fig. 1. The Jarlskog invariant ${\cal J}$ may
vary from 0 to 0.057 in the obtained parameter space.

\vspace{0.3cm}

\framebox{\large\bf 4} ~ To summarize, we have written out the
generic Dirac neutrino mass operator which possesses the FL
symmetry and pointed out that its corresponding neutrino mass
matrix is actually asymmetric. After introducing a perturbative
term to break the FL symmetry, we have calculated the neutrino
mass eigenvalues and flavor mixing angles. We find that the
resultant neutrino mixing pattern is nearly tri-bimaximal.
Imposing the Hermitian condition on the neutrino mass matrix, we
have shown that the simplified ansatz is consistent with current
experimental data and favors the normal neutrino mass hierarchy.

This work is a simple but useful generalization of the original FL
symmetry for Dirac neutrinos. Such a generic FL symmetry can be
applied to the quark sector to obtain generic (or Hermitian) quark
mass matrices. But it will have no influence on the neutrino mass
matrix if massive neutrinos are Majorana particles, because a
Majorana neutrino mass matrix must always be symmetric.

In conclusion, the FL symmetry and its breaking mechanism may have
a wealth of implications in neutrino phenomenology. The physics
behind this interesting flavor symmetry remains unclear to us and
deserves a further study, no matter whether massive neutrinos are
Dirac fermions or Majorana fermions.

\vspace{0.35cm}

One of us (Z.Z.X.) likes to thank G.J. Ding for having asked a
correct question. We are also grateful to S. Zhou for useful
discussions. This work was supported in part by the National Natural
Science Foundation of China.

\newpage

\begin{figure}
\begin{center}
\vspace{2cm}
\includegraphics[bbllx=6.5cm, bblly=3.0cm, bburx=14.0cm, bbury=15.2cm,%
width=10cm, height=16cm, angle=0, clip=0]{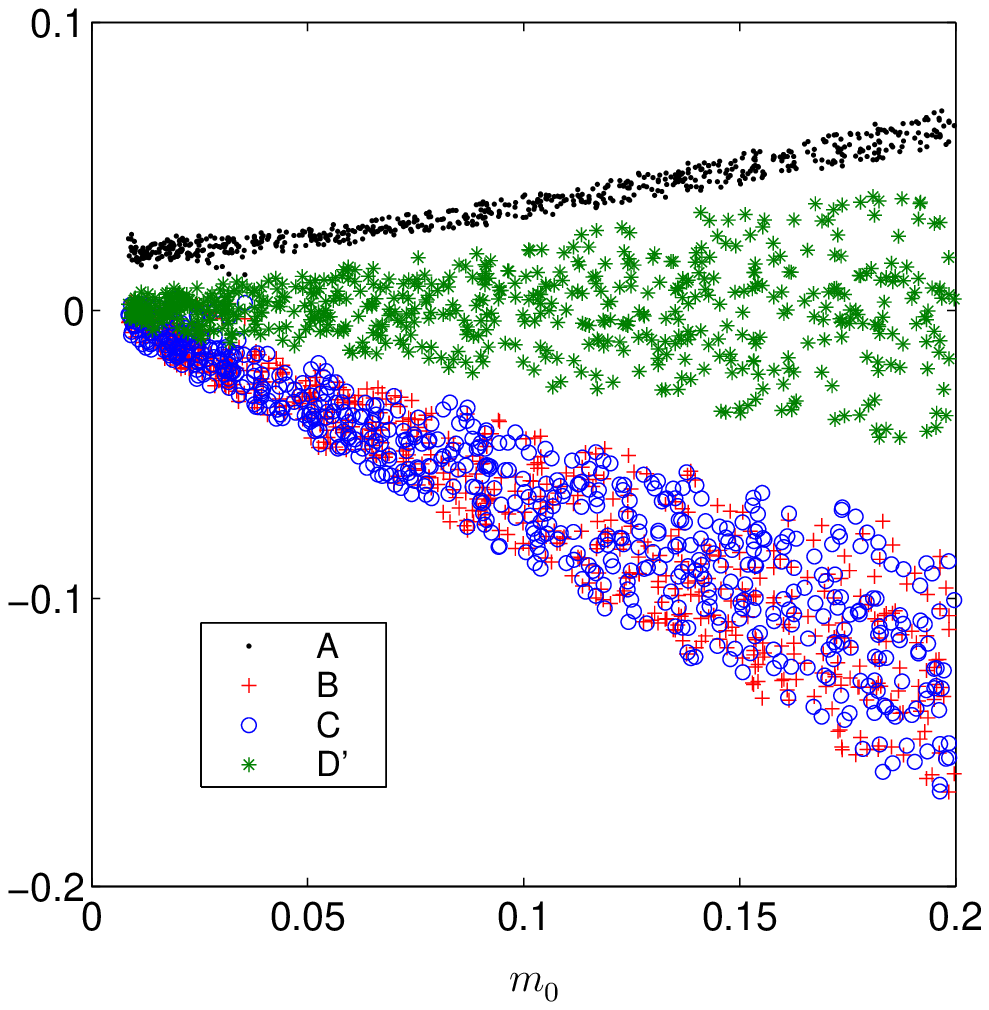}
\vspace{-1.5cm}\caption{The parameter space of $A$, $B$, $C$ and
$D'$ versus $m^{}_{0}$ (all of them in unit of eV) in the scenario
of Hermitian $M^{}_{\nu}$, where only the normal neutrino mass
hierarchy (i.e., $\Delta m^2_{32} >0$) is allowed.}
\end{center}
\end{figure}


\begin{thebibliography}{99}

\bibitem{SNO} SNO Collaboration, Q. R. Ahmad {\it et al.},
Phys. Rev. Lett. {\bf 89}, 011301 (2002).

\bibitem{SK} For a review, see: C. K. Jung {\it et al.},
Ann. Rev. Nucl. Part. Sci. {\bf 51}, 451 (2001).

\bibitem{KM} KamLAND Collaboration, K. Eguchi {\it et al.},
Phys. Rev. Lett. {\bf 90}, 021802 (2003).

\bibitem{K2K} K2K Collaboration, M. H. Ahn {\it et al.},
Phys. Rev. Lett. {\bf 90}, 041801 (2003).

\bibitem{MNS} Z. Maki, M. Nakagawa and S. Sakata, Prog. Theor. Phys. {\bf 28} 870 (1962).

\bibitem{PDG} Particle Data Group, C. Amsler {\it et al.}, Phys. Lett. B {\bf 667}, 1 (2008).

\bibitem{Fogli} G. L. Fogli {\it et al.}, arXiv:0805.2517 [hep-ph];
T. Schwetz, M. Tortola and J. W. F. Valle, arXiv:0808.2016 [hep-ph].

\bibitem{Review} For recent reviews with extensive references,
see: H. Fritzsch and Z. Z. Xing, Prog. Part. Nucl. Phys. {\bf 45}, 1
(2000); Altarelli and F. Feruglio, New J. Phys. {\bf 6}, 106 (2004);
R. N. Mohapatra and A. Yu. Smirnov, Ann. Rev. Nucl. Part. Sci. {\bf
56}, 569 (2006); A. Strumia and F. Vissani, hep-ph/0606054.

\bibitem{FL} R. Friedberg and T. D. Lee, High Energy Phys. Nucl.
Phys. {\bf 30}, 591 (2006).

\bibitem{XZZ} Z. Z. Xing, H. Zhang and S. Zhou, Phys. Lett. B {\bf 641}, 189
(2006); arXiv:0712.2611 [hep-ph].

\bibitem{LX} S. Luo and Z. Z. Xing, Phys. Lett. B {\bf 646}, 242
(2007).

\bibitem{Xing07} Z. Z. Xing, Int. J. Mod. Phys. E {\bf 16}, 1361 (2007).

\bibitem{Jarlskog} C. Jarlskog, Phys. Rev. D {\bf 77}, 073002
(2008).

\bibitem{Chao} W. Chao, S. Luo and Z. Z. Xing,
Phys. Lett. B {\bf 659}, 281 (2008).

\bibitem{FL2} R. Friedberg and T. D. Lee, A
nnals Phys. {\bf 323}, 1087 (2008); Annals Phys. {\bf 323}, 1677
(2008).

\bibitem{Ren} P. Ren, arXiv:0801.0501 [hep-ph].

\bibitem{Liao} C. S. Huang, T. J. Li, W. Liao and S. H. Zhu, arXiv:0803.4124 [hep-ph].

\bibitem{Dirac} See, e.g., R. N. Mohapatra and J. W. F. Valle,
Phys. Rev. D {\bf 34}, 1642 (1986); N. Arkani-Hamed {\it et al.},
Phys. Rev. D {\bf 64}, 115011 (2001); F. Borzumati and Y. Nomura,
Phys. Rev. D {\bf 64}, 053005 (2001); R. Kitano, Phys. Lett. B
{\bf 539}, 102 (2002); R. Arnowitt, B. Dutta and B. Hu, Nucl.
Phys. B {\bf 682}, 347 (2004); S. Abel, A. Dedes and K. Tamvakis,
Phys. Rev. D {\bf 71}, 033003 (2005); P. Q. Hung, Nucl. Phys. B
{\bf 720}, 89 (2005); P. Ko, T. Kobayashi and J. H. Park, Phys.
Rev. D {\bf 71}, 095010 (2005); J. Giedt {\it et al.}, Phys. Rev.
D {\bf 71}, 115013 (2005); C. Hagedorn and W. Rodejohann, JHEP
{\bf 0507}, 034 (2005); S. Antusch, O. J. Eyton-Williams and S. F.
King, JHEP {\bf 0508}, 103 (2005); M. Lindner, M. Ratz and M. A.
Schmidt, JHEP {\bf 0509}, 081 (2005); Z. Z. Xing and H. Zhang,
Commun. Theor. Phys. {\bf 48}, 525 (2007).

\bibitem{TB} P. F. Harrison, D. H. Perkins, and W. G. Scott, Phys.
Lett. B {\bf 530}, 167 (2002); Z. Z. Xing, Phys. Lett. B {\bf 533},
85 (2002); P. F. Harrison and W. G. Scott, Phys. Lett. B {\bf 535},
163 (2002).

\bibitem{J} C. Jarlskog, Phys. Rev. Lett. {\bf 55}, 1039 (1985);
H. Fritzsch and Z. Z. Xing, Nucl. Phys. B {\bf 556}, 49 (1999).

\bibitem{WMAP} WMAP Collaboration, E. Komatsu {\it et al.},
arXiv:0803.0547 [astro-ph], where $\sum m^{}_i < 0.61$ eV ($95\%$
C.L.) has been presented.

\end{thebibliography}
\end{document}